\documentstyle[epsfig]{aipproc}

\begin{document}
\title{Evidence for GeV emission from the Galactic Center Fountain}

\author{D. H. Hartmann$^*$, D. D. Dixon$^{**}$, E. D. 
Kolaczyk$^{\dagger}$,
J. Samimi$^{\dagger\dagger}$}
\address{$^*$Department of Physics and Astronomy\\
		Clemson University, Clemson, SC 29634\\
	$^{**}$Institute of Geophysics and Planetary Physics\\
		University of California Riverside, CA 92521\\
	$^{\dagger}$Department of Statistics\\
		University of Chicago, Chicago IL\\
	$^{\dagger\dagger}$Sharif University\\
		Tehran, Iran
}

\maketitle

\begin{abstract}
The region near the Galactic center may have experienced recurrent episodes of 
injection of energy in excess of $\sim$ 10$^{55}$ ergs due to repeated 
starbursts
involving more than $\sim$ 10$^4$ supernovae.
This hypothesis can be tested by measurements of $\gamma$-ray lines produced by 
the decay of radioactive isotopes and positron annihilation, or by searches for 
pulsars produced during starbursts. Recent OSSE observations of 511 keV emission
extending above the Galactic center led to the suggestion of a starburst
driven fountain from the Galactic center~\cite{DS97}. We present
EGRET observations that might support this picture.
\end{abstract}

\section*{The Galactic Center Fountain}

The center of the Milky Way may have experienced a series of explosive events
~\cite{Oort}~\cite{Loose}~\cite{Sofue89}.
Large scale X-ray structures, such as the north polar spur, might be explained
as propagating shocks induced by these explosions ~\cite{Sofue94}, although 
other
arguments suggest the north polar spur is a rather local feature.
To understand the angular distribution of the X-ray features, the shock
model requires an impulsive energy release of $\sim$ 3 $\times$ 10$^{56}$ ergs
about $\sim$ 1.5 $\times$ 10$^7$ yrs ago.  A massive black hole at the Galactic
center could have released this energy, but an alternative scenario is the
energy deposition from a large number of supernovae. 

A typical Type II supernova injects about 1--2 $\times$ 10$^{51}$ ergs of
kinetic energy into the ISM, and contributions from pre-supernova winds may
double this amount.
Thus, a total of $\sim$ 10$^5$ supernovae is needed to attain a total
energy of 3 $\times$ 10$^{56}$ ergs. The duration of the starburst should be
$<$ 10$^6$ yrs in order to yield a propagating shock that matches the observed
X-ray features~\cite{Sofue94}.  Activities observed near the Galactic center are
manifest on various spatial scales, with perhaps the most dominant feature
being the expanding molecular ring.  At a Galactocentric distance of $\sim$ 200
pc, the expanding molecular ring might contain as much as 10$^{55}$ ergs of
kinetic energy ~\cite{Morris}.  Enhanced 6.7 keV line emission was detected by
the GINGA satellite at a distance compatible with being associated with the
ring structure ~\cite{Koyama89}~\cite{Koyama90}. If this line emission
originates from a hot and tenuous plasma, then the X-ray observations suggest
$\sim$ 10$^{54}$ ergs of thermal energy were injected less than $\sim$ 10$^6$
yrs ago ~\cite{Yamauchi90}. 

On smaller scales, the Galactic superbubble G359.1--0.5 suggests that a
starburst of $\sim$ 10$^{2-3}$ supernovae may have occured within the past few
million years ~\cite{Uchida92}.  Observations of stellar populations
within $\sim$ 1 pc of the Galactic center argue in favor of starburst models
involving $\sim$ 4 $\times$ 10$^5$ M$_\odot$ of gas (implying $\sim$ 10$^{2-3}$
supernovae) between 5 -- 9 Myr ago ~\cite{Tamblyn}.  Emission features
from He I surveys of the central region also implies that at least several tens
of massive stars were born within a few parsecs of the center in the last
$\sim$ 10$^6$ yrs ~\cite{Krabbe91}~\cite{Rieke94}.

Other arguments support more extensive or intense starburst episodes.  The
total mass interior to 1 pc exceeds 10$^6$ M$_\odot$ ~\cite{Genzel94}.
If a significant fraction of this total mass is due to mass
segregation of compact stellar remnants initially formed within the inner
10--100 pc, then the associated number of neutron stars could exceed $\sim$
10$^6$, for a Salpeter IMF ~\cite{Morris}. However, the velocity that neutron
stars are apparently born with may allow most of them to escape the central
region.  Thus, neutron stars would not contribute significantly to the mass
interior to 1 pc.  If the neutron stars were not produced in steady state but
in a series of starbursts, one might consider the production of $\sim$ 10$^3$
neutron stars in bursts separated $\sim$ 10$^7$ yrs.  Such starbursts would
also inject over 10$^{54}$ ergs of kinetic energy into the ISM.  As noted
above, however, the expanding molecular ring imply that the last energy
deposition was a factor 10$-$100 larger ~\cite{Sofue89}. Hydrodynamic 
simulations
of gas--star systems near galactic centers suggest that starbursts which
produce $>$ 10$^5$ supernovae could occur quasi-periodically every $\sim$
10$^8$ yrs ~\cite{Loose}. Bursts of this magnitude would be
expected to severely influence the gas dynamics near the center, and (to a
lesser extent) the disk and the halo through the influence of the propagating
shock wave.  

New evidence for a recent starburst in the inner Galaxy comes from the 511 keV
mapping by OSSE~\cite{Cheng}~\cite{Purcell}. The global map can be decomposed 
into two components, a disk and a bulge. In addition, the data require a ``hot
spot'' at l $\sim$ $-$4$^{\rm o}$ and b $\sim$ 7$^{\rm o}$. This positive
latitude enhancement was interpreted by Dermer $\&$ Skibo~\cite{DS97} as the
result of a recent starburst ($\sim$ 10$^6$ yrs ago) involving $\sim$ 10$^4$
supernovae. The resulting positrons lose energy and annihilate as they are 
convected upward with the gas flow. In this picture one also expects the
coproduction of $^{26}$Al (visible on a timescale of 10$^6$ yrs) and cosmic
rays (CRs). Shocks would also produce a non-thermal population of electrons
which might produce a radio afterglow. In fact, a 4 kpc long jet-like radio
feature emanating from the Galactic center region has been detected at 408 MHz
~\cite{SRR}, and is commonly known as the Galactic Center Spur (GCS). If this
$\sim$ 200 pc wide chimney indeed convects radiactive debris and CRs into
the halo, we might also expect some emission in the GeV regime due to 
interactions
of the CRs with the gas in the chimney. 

\section*{Detection of GeV Emission}

The data analyzed are coadded EGRET observations through VP 429.0,
selecting only events within 30 degrees of the detector zenith.
The analysis method is a 2D variant~\cite{dixon1} of the TIPSH
algorithm for denoising Poisson data. 
In the particularly TIPSH method
employed here, we specify a null hypothesis, consisting of a predicted
distribution of counts/pixel in the data set.  Here, we have used a
hypothesis consisting of the predicted Galactic~\cite{Hunter} and
extragalactic~\cite{kumar} diffuse emission.  TIPSH works by comparing
the Haar wavelet coefficients of the data with the distribution of
coefficients implied by a Poisson distribution whose pixel means 
(and variances) are described by the null hypothesis.  Those coefficents
which fall below some user prescribed significance cutoff are considered
statistically consistent with the null hypothesis, and discarded.  At the
end, the non-zero wavelet coefficients describe the portion of the data
which is ``different'' than the null hypothesis, within the statistics
of the observations, along with some (preferably small) number of false
detections (non-zero coefficients due solely to noise).  For the analysis
described here, we selected our significance threshhold such that the
error rate was about 2\%.

For the region under discussion, the denoised residual (significant
differences w.r.t. null hypothesis) for the 4-10 GeV band
is shown in Figure 1, overplotted
on filled contours showing the 511 keV model~\cite{Purcell}. The
Galactic plane flux has been truncated to show the feature of interest, which
is an apparent northward extension of the Galactic emission, at
$1^\circ-2^\circ$ longitude, extending up to about $15^\circ$ latitude
as seen in this plot.  Though we have not yet derived a spectrum for
this feature, similar analysis in the 1-2 GeV band shows no evidence
for a similar feature.  In the 2-4 GeV band, there does appear to be
enhanced emission in this region.  However, it is confused with other
nearby emission, and is not nearly so distinct.  A reasonable conclusion
is that the spectrum of this feature is fairly hard, and distinctly
different than that of Galactic cosmic-ray induced emission.  Further
model fitting is required to verify this. Note that this feature coincides
with the ``jet'' seen in the radio band.  Also in Figure 1 are
filled contours showing the most recent OSSE 511 keV model 
fit~\cite{Purcell}. Though there is an apparent latitude offset in
the positive latitude feature, it can be shown
that due to a strong exposure-related systematic in the OSSE 
observations~\cite{dixon2}, 
a 511 keV feature corresponding to the EGRET/408MHz
``jet'' would be consistent with the observed 511 keV maps.

A key question is the reality of the observed features, since
when using a non-parametric estimation scheme such as TIPSH, one is
always concerned with artifacts.  Unfortunately, it is usually difficult
(if not impossible) to assign a quantitative ``significance'' to a feature
in a non-parametric estimate.  At the time of this writing, we have not
yet devised a method to accomplish this.  An alternative approach would
be to perform some sort of model fitting, but this has its own set of
pitfalls, and must be accomplished with care.  Future analyses will
attempt to address this, hopefully using physically motivated models.


\begin{figure}[t]
\centerline{\epsfig{file=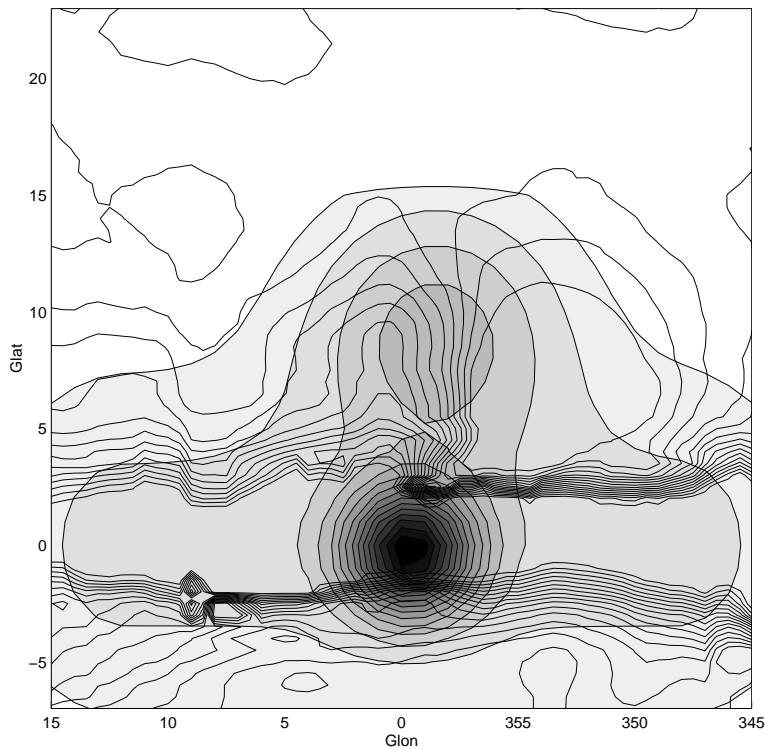,width=4.0in}}
\caption[fig1]{Flux contours from EGRET counts in the 4-10 GeV band. 
The jet-like feature is
reasonably well aligned with the galactic center spur seen in 408 MHz maps.
The filled contours represent the 511 keV model fit described in
\cite{Purcell}.  The apparent offset between the 511 keV and GeV features
is potentially due to exposure systematics in the OSSE observations.}
\end{figure}

\section*{Conclusions}

Whether driven by bursts of star formation or processes that occur near a
massive black hole, the numerous activities going on near the Galactic center
are hidden, for the most part, from optical observations. In the gamma-ray
band evidence for starburst activity is harder to hide. Hartmann, Timmes,
and Diehl~\cite{Hartmann96}
discussed the possibility that the production of $^{26}$Al
in supernovae may lead to a detectable afterglow at 1.809 MeV, and 
Hartmann~\cite{Hartmann95}
suggested that such starbursts might be detectable through an excess
of radio pulsars.  The detection of 1.8 MeV emission from a galactic center
starburst might be accomplished by the INTEGRAL mission. The recent
OSSE observations of 511 keV emission above the galactic center~\cite{Cheng}
~\cite{Purcell} were interpreted as the
result of a major galactic center starburst driving a positron fountain
into the halo ~\cite{DS97}. We present EGRET GeV observations 
that perhaps support this picture, but new gamma-ray missions such as GLAST will 
be required to verify these observations. The Galactic center is one of the 
most dynamical regions of our Galaxy, and high energy gamma rays may be the
best tool for studying its starburst history.

\medskip
This work was supported by NASA Grant NAG5-3666 (DDD). DH acknowledges support
from the Compton GI program.

\end{document}